\documentclass[aps,prl,reprint,showpacs]{revtex4-1}

\draft 

\usepackage{graphicx}
\usepackage{dcolumn}
\usepackage{bm}
\usepackage{xcolor}
\usepackage{changebar}

\definecolor{deep_pink}{RGB}{255,20,147}
\newcommand{\corrected}[1]{\cbstart\correctedfragile{#1}\cbend{}}
\newcommand{\correctedfragile}[1]{{\color{black}{#1}}{}}

\begin{document}

\title{Plasmonic Nanoantennas for Efficient Control of Polarization-Entangled Photon Pairs} 

\author{Ivan S. Maksymov}
\author{Andrey E. Miroshnichenko}
\author{Yuri S. Kivshar}

\affiliation{Nonlinear Physics Centre and Centre for Ultrahigh Bandwidth Devices for Optical Systems (CUDOS), Research School of Physics and Engineering, The Australian National University, Canberra, ACT 0200, Australia. E-mail: mis124@physics.anu.edu.au}

\date{\today}

\begin{abstract}

\corrected{We suggest a novel source of polarization-entangled photon pairs based on a cross-shaped plasmonic nanoantenna driven by a single quantum dot. The integration of the nanoantenna with a metal mirror overcomes the fundamental tradeoff between the spontaneous emission (SE) enhancement and the extraction efficiency typical of microcavity and nanowire-based architectures. With a {\color{red}very}-high extraction efficiency of entangled photons ($\approx90\%$) at $1.55$ $\mu$m and large SE enhancement ($\approx90$) over a broad $330$ nm spectral range, the proposed design will pave the way toward reliable integrated sources of nonclassical light.} 

\end{abstract}


\maketitle 

Quantum communication systems and computers can achieve unparalleled levels of security and productivity by encoding information using polarization-entangled photon pairs \cite{bou00}. Photonic chips capable of both generating polarization-entangled photon pairs and performing logic quantum operations \cite{hau12, *sol12, *hor12, *sha12} are a promising technique \corrected{for developing practical} quantum communication and computers. \corrected{It would be} impossible without compact, integrable and scalable sources of polarization-entangled photon pairs combining broad bandwidth, large efficiency and high fidelity \cite{bou00, hen08}.

\begin{figure}
\includegraphics[width=6.5cm]{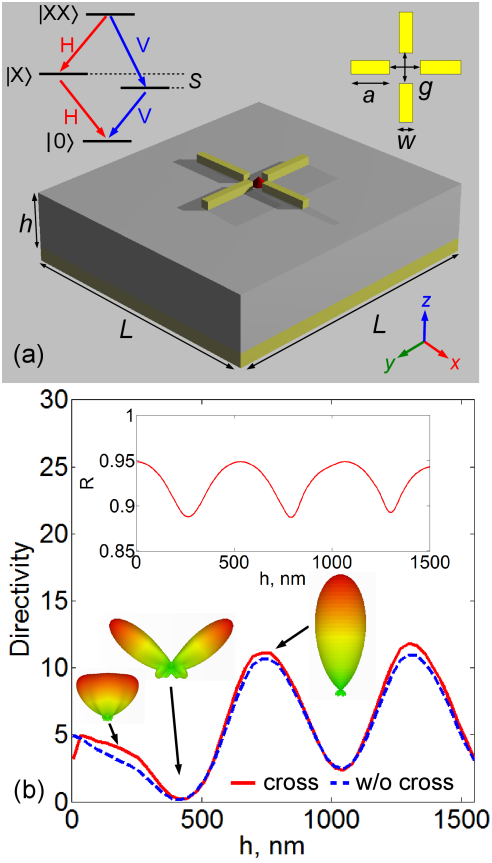}
\caption{(a) Schematic of the cross-shaped plasmonic nanoantenna on a pedestal glass substrate of height $h$ and side size $L$ with an underlying gold mirror. The mirror is on the top of an infinitely wide glass bulk. The left inset sketches the radiative cascade of a single QD placed in the common gap of the cross (the red arrow in the main panel). The right inset defines the dimensions of the cross. (b) The directivity of the far-field radiation from the QD at $\lambda=1550$ nm, $L=2000$ nm and gold mirror height of $150$ nm with the cross (red solid line) and without the cross (blue dashed line). The characteristic far-field pattern are shown for three values of $h$. The inset shows the reflectivity at normal incidence of an infinitely wide glass layer of thickness $h$ placed over a gold bulk.}
\end{figure}

{\color{red}Single semiconductor quantum dots (QDs) can be used as a triggered source of polarization-entangled photon pairs \cite{[{For a review see, e.g., }][{, and references therein for a discussion of experimental demonstration of entangled photon pairs with single QDs.}]shi07}.} As shown in the inset in Fig. $1$(a), the radiative decay of the biexciton state (\textit{XX}) in a QD emits a pair of photons, with the polarization determined by the spin of the intermediate exciton state (\textit{X}). In an ideal QD with degenerate \textit{X} states (the energy splitting $S=0$ \cite{shi07}), the polarization of the \textit{XX} photon is predicted to be entangled with that of \textit{X} photon, forming the state $(|H_{XX}H_{X}\rangle+|V_{XX}V_{X}\rangle)/\sqrt{2}$, where \textit{H} and \textit{V} denote the polarization of the \textit{XX} and \textit{X} photons.

In practice, \corrected{a strong scattering of emitted entangled photons over a broad angular spectrum and the resulting far-field divergence severely limit the extraction efficiency. Moreover,} the degree of entanglement is still limited by QD dephasing and residual excitonic fine-structure splitting \cite{hen08}. The use of cavity effect \cite{shi07,lar09,dou10} improves this situation dramatically: the emission is directed into a single mode and the emission lifetime is reduced due to the SE enhancement effect, which makes the source more deterministic and reliable. Conversely, one can avoid cavity effect and utilize photonic crystal waveguides \cite{lec07, *man07, *lun08} or semiconductor nanowires \cite{cla10, *bab10, *rei12}. While these architectures may offer high multiphoton probability suppression \cite{lun08} and very high extraction efficiency through a microscope objective \cite{cla10, *bab10, *rei12}, they provide a low SE enhancement insufficient for the restoration of polarization entanglement \cite{hen08, lar08}. Large SE rates can be achieved {\color{red} using various metallic nanostructures \cite{ang06,*kin09,*sch09,*sor11,*cho11_1,*rus12}, including metal-coated semiconductor nanowires \corrected{\cite{cho11,*gan12,mak10}}}. However, the extraction efficiency of entangled photons in such structures is compromised by impedance mismatch between the confined nanowire modes and freely propagating light \cite{mak10}.

Recent advances in nanofabrication have demonstrated that optical nanoantennas \cite{novotny_review}, such as, e.g., plasmonic Yagi-Uda\corrected{-type architectures driven by a single emitter \cite{koe09,*cur10,*est10,*mak11}}, may have much larger SE enhancement as compared with nanocavities and nanowires. Moreover, in general, a nanoantenna is an intrinsically good transformer of localized energy to freely propagating light \cite{novotny_review}, which favours high extraction efficiency of the emitted photons into a cone with a narrow opening angle given by the numerical aperture of the detection optics \cite{lee11, *che11}. However, the use of many nanoantenna architectures for the emission of polarization-entangled photon pairs is challenging because they provide an increased SE enhancement for one photon polarization \cite{car06}, which leads to the generation of nonmaximally entangled states \cite{hen08}.

In this Letter, we suggest a novel QD-based source of polarization-entangled photon pairs benefiting from a broadband SE rate enhancement and, to our best knowledge, {\color{red}very}-high extraction efficiency of photon pairs. This is achieved with a cross-shaped plasmonic nanoantenna {\color{red} inspired by radio-frequency cross antennas \cite{bal05} recently adopted in optics (see, e.g., \cite{bia09}). In contrast to Ref.~\cite{bia09}, our cross structure is integrated with a metal mirror similar to that demonstrated in the literature \cite{fri09, *ahm11}.} The mirror plays a crucial role in the performance of the device since it forms a sharp far-field emission pattern and facilitates increased extraction efficiency. We also demonstrate that a pure entangled photon emission with a high fidelity is \corrected{still possible for the misaligned QDs from the centre} of the cross. The proposed design is consequently a major contributor to the development of efficient sources of nonclassical light for photonic quantum technologies.

The design and interpretation were performed using CST Microwave Studio implementing a finite integration technique, which was tested on analytically solvable models \cite{car06}. Except when otherwise specified, the emitter is modelled as an in-plane electric dipole polarized horizontally [along the \textit{x}-axis in Fig. $1$(a)] and located at the center of the cross. Since the cross displays symmetry, the results obtained with the horizontal dipole can be generalized for the case of the vertically polarized dipole [along the \textit{y}-axis in Fig. $1$(a)] representing photons of \textit{V}-polarization. 

The cross consists of four identical gold nanorods with a common feed gap [Fig. $1$(a), main panel]. It sits on a pedestal glass substrate of height $h$ and a side size $L$ with an underlying gold mirror. The mirror resides on the top of an infinitely wide glass bulk. For the optimal performance around $\lambda=1550$ nm the length of the nanorods is chosen to be $a=250$ nm and their width and height are $w=30$ nm. The central gap between the nanorods is $g=75$ nm in both the \textit{x}- and \textit{y}-directions. The frequency-dependent refractive index of gold is taken from Ref.~\cite{palik}; that of glass is chosen as $1.5$. 

\begin{figure}
\includegraphics[width=5.5cm]{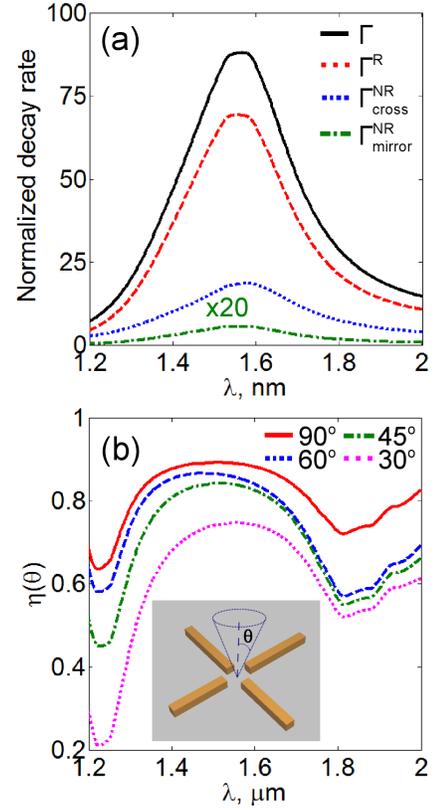}
\caption{Spectral performance of the nanoantenna. (a) Total $\Gamma$, radiative $\Gamma^{\rm{R}}$ and nonradiative $\Gamma^{\rm{NR}}$ normalized decay rates due to absorption by the cross and the mirror. $\Gamma^{\rm{NR}}_{\rm{mirror}}$ is zoomed for illustration's sake. (b) Extraction efficiency for several opening angles $\theta$. Owing to the symmetry of the cross, the same result holds for both \textit{H}- and \textit{V}-polarized dipoles.}
\end{figure}

The choice of the area of the pedestal substrate is of crucial importance for achieving a directive far-field pattern of the nanoantenna. It is known that the maximum directivity $D_{\rm{max}}$ obtainable from an antenna is proportional to the physical aperture $A_{\rm{p}}$ \cite{kraus}. For the cross nanoantenna in Fig. $1$(a) $A_{\rm{p}}$ is determined by the area of the pedestal substrate. In auxiliary simulations we observe that the excitation of the nanoantenna sitting on a large area substrate ($L>>\lambda$) results in launching surface plasmons propagating along the mirror surface. It enables plasmons contributing to the emission pattern as highly undesirable side lobes. The formation of side lobes is suppressed by using a pedestal substrate with a mirror of a smaller size $L=2000$ nm. 

In order to gain more insight into the directivity of the nanoantenna, in Fig. $1$(b) we compare $D_{\rm{max}}$ of the structures with and without the cross and observe that $D_{\rm{max}}$ depends on the distance $h$ between the emitter and the mirror. The cross plays a little role in the formation of the emission pattern. By investigating the reflectivity $R$ of the infinitely wide glass layer on the top of the gold bulk as a function of $h$ [inset in Fig. $1$(b)] we reveal that a directive emission pattern is formed solely by the mirror under condition that the reflection from the covering glass is minimum. This condition fulfils at, e.g., $h=750$ nm corresponding to the second minimum of $R$ and, therefore, in what follows we consider this value only. The first minimum of $R$ occurs when the emitter is close to the mirror, which incurs increased absorption losses and undesired broad emission pattern. 

By definition, the directivity does not manifest the enhancement of the SE rate provided by the nanoantenna. We therefore consider the normalized spontaneous decay rate that, in the weak coupling regime, is related to the rate of energy dissipation experienced by a classical electric dipole: $\Gamma/\Gamma_{\rm{0}}=P/P_{\rm{0}}$, where $P$ is the power emitted by the dipole in the presence of the nanoantenna and $P_{\rm{0}}$ is the power of the same dipole in free space \cite{car06}. We calculate the normalized radiative decay rate $\Gamma^{\rm{R}}/\Gamma_{\rm{0}}=P_{\rm{R}}/P_{\rm{0}}$, being $P_{\rm{R}}$ the far-field radiated power. As an increase in $\Gamma^{\rm{R}}$ is typically concomitant with an unwanted and usually strong increase in the nonradiative decay rate $\Gamma^{\rm{NR}}$, we also calculate $\Gamma^{\rm{NR}}/\Gamma_{\rm{0}}=P_{\rm{abs}}/P_{\rm{0}}$, where $P_{\rm{abs}}$ is the power absorbed inside the cross and the mirror. Owing to energy conservation the total decay rate is $\Gamma=\Gamma^{\rm{R}}+\Gamma^{\rm{NR}}$.

Fig. $2$ shows the spectral characteristics of the nanoantenna. The peak value of $\Gamma$ in Fig. $2$(a) is $\approx90$ at the wavelength of $\lambda=1550$ nm, which is $6$ times larger than that of a recently reported metal-coated nanowire sources of entangled photon pairs \cite{mak10}. The peak value of $\Gamma^{\rm{R}}$ reaches $70$ giving rise to a radiation efficiency of $\frac{\Gamma^{\rm{R}}}{\Gamma^{\rm{R}}+\Gamma^{\rm{NR}}}\approx0.79$ at $\lambda=1550$ nm. Furthermore, the cross nanoantenna is broadband in frequency having a remarkable $330$ nm bandwidth at the full-width at half-maximum of $\Gamma$, which is $5$ times larger than the bandwidth of semiconductor and metal-coated nanowire-based sources of single photons \cite{cla10, *bab10, *rei12, mak10}. The extraction efficiency $\eta$ in Fig. $2$(b), defined as a fraction of emitted photons that are collected above the top of the cross for the opening angle $\theta$ [inset in Fig. $2$(b)], is greater than $70\%$ for an opening angle as small as $\theta=30^{\rm{o}}$ in a broad $250$ nm range. In overall, $\eta$ observed is comparable with that of state-of-the-art photonic nanowire single-photon sources known to boast very high extraction efficiency \cite{cla10, *bab10, *rei12}. It is very close to that of recently demonstrated planar dielectric nanoantennas for single-photon emission \cite{lee11, *che11}, which, however, do not provide any SE enhancement.

\begin{figure}
\includegraphics[width=7cm]{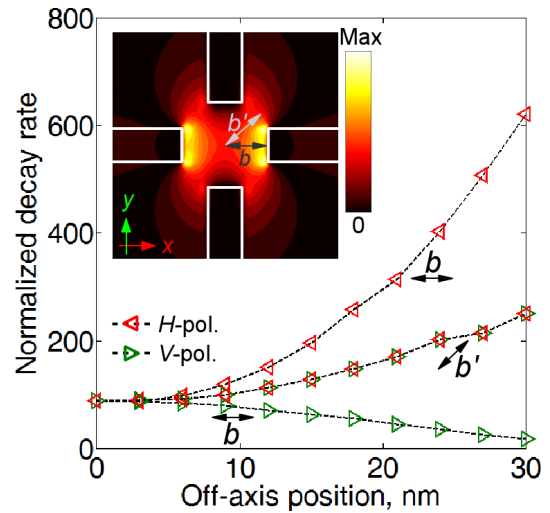}
\caption{Influence of the axial $b$ and diagonal $b'$ off-axis dipole displacement at $\lambda=1550$ nm. Red left-pointing and green right-pointing triangles denote $\Gamma$ of the \textit{H}- and \textit{V}-dipoles. Dashed lines are the guide to the eye. Inset: $|E_{\rm{x}}|$ field profile. By virtue of symmetry, $|E_{\rm{y}}|$ equals to $|E_{\rm{x}}|$ rotated by $90^{\rm{o}}$.}
\end{figure}

Thus far we have assumed that the nanoantenna is excited with a QD precisely placed at the center. In this case the SE  enhancement and far-field emission pattern underwent by the \textit{H}- and  \textit{V}-dipole are identical due to the mode degeneracy. In practice, however, the QD will be not exactly centered. Fig. $3$ shows $\Gamma$ of the \textit{H}- and \textit{V}-polarized dipoles as a function of the axial ($b$) and diagonal ($b'$) displacement from the central position (inset in Fig. $3$). $\Gamma$ of the \textit{H}-dipole increases rapidly as the dipole approaches one of the cross arms. Since the total power emitted by the dipole is proportional to the electric field \textbf{E} at the dipole's origin \cite{jackson}, the increase in $\Gamma$ is due to an increase in $E_{\rm{x}}$ in the direction of the cross arm (inset in Fig. $3$). A rapid decrease in $\Gamma$ for the \textit{V}-dipole is due to a rapid decrease in $E_{\rm{y}}$. This lift of degeneracy results in an increased SE enhancement for one recombination path and inevitably leads to the generation of nonmaximally entangled states \cite{hen08}. As for the \textit{H}- and \textit{V}-dipoles displaced along the diagonal, we find that their $\Gamma$ changes simultaneously because of the same variations of $E_{\rm{x}}$ and $E_{\rm{y}}$ along the diagonal direction. 

\corrected{The asymmetry of $\Gamma$ influences the degree of entanglement, which can be quantified by the fidelity $\mathcal{F}$ of entangled photons pairs \cite{lar09, dou10}. $\mathcal{F}$ measures the similarity of the emitted state to the maximally entangled Bell state $|\psi^{+}\rangle=(|H_{XX}H_{X}\rangle+|V_{XX}V_{X}\rangle)/\sqrt{2}$. Under the assumption that the anisotropic spin exchange energy of the QD is negligible, we estimate the density matrix of the photon pairs in the polarization basis as} \cite{lar09,[{A. Beveratos (personal communication). Eq. (C.4) in Ref.~\onlinecite{lar09} contains misprints. We use the correct equation}][{}]dummy}
\begin{eqnarray}
\rho = \frac{1}{2(1+\delta \Gamma^2)}
\left(
\begin{array}{cccc}
(\delta \Gamma + 1)^2&0&0&(1-\delta \Gamma^2)^{2} \\
0&0&0&0 \\
0&0&0&0 \\
(1-\delta \Gamma^2)^{2}&0&0&(\delta \Gamma - 1)^2
\end{array}\right)
\label{eq:two},
\end{eqnarray}
\noindent \corrected{being $\delta \Gamma = (\Gamma_{\rm{H}} - \Gamma_{\rm{V}})/(\Gamma_{\rm{H}} + \Gamma_{\rm{V}})$ the relative difference of $\Gamma$ for the \textit{H}- and \textit{V}-dipoles}. In Fig. $4$ we plot $\mathcal{F}$ calculated as \corrected{Tr($\rho|\psi^{+}\rangle\langle\psi^{+}|)$}, where Tr(.) is the trace and $\rho$ is the density matrix of the entangled photons emitted by a QD displaced from the central position. We observe $\mathcal{F}>0.85$ required to violate the original Bell inequality \cite{bou00} for axial displacements of up to $\pm12$ nm. For the diagonal displacement \corrected{$\mathcal{F}=1$} because $\Gamma_{\rm{H}}=\Gamma_{\rm{V}}$ (see Fig. $3$). 

{\color{red} Precise positioning of a single QD within the gap of the cross structure is challenging but possible thanks to the recent advances in nanofabrication technologies. For example, strong coupling of a single QD with a cavity architecture was achieved in Ref.~\cite{tho09} using an all-optical scheme ensuring a sub-10 nm accuracy. Topdown fabrication techniques were used in Ref.~\cite{dre12} to achieve positioning of individual quantum emitters relative to plasmonic nanoantennas with an accuracy better than 10 nm. More importantly, the combination of the cross nanoantenna with the gold mirror allows relaxing the requirement for precise positioning of the QD by increasing the gap of the cross. It is possible because $\eta$ is practically independent of the presence of the cross structure [see Fig. $1$(b)]. Simulations for the cross gap of $g=150$ nm reveal (Fig. $4$) that $\mathcal{F}>0.85$ for axial displacements of $\pm24$ nm implying that more mature QD alignment procedures with an accuracy of better than 30 nm can be used \cite{pfe12}. Eventually, $\Gamma$ decreases as increases the gap. For $g=150$ nm we obtain  $\Gamma=12$ and $\Gamma^{\rm{R}}=10$ at $\lambda=1550$ nm [Supplementary Material] with the same extraction efficiency $\eta$ as for $g=75$ nm. We note that $\Gamma^{\rm{R}}=10$ suffices for the emission of entangled photons \cite{lar08}.}

\begin{figure}
\includegraphics[width=5.5cm]{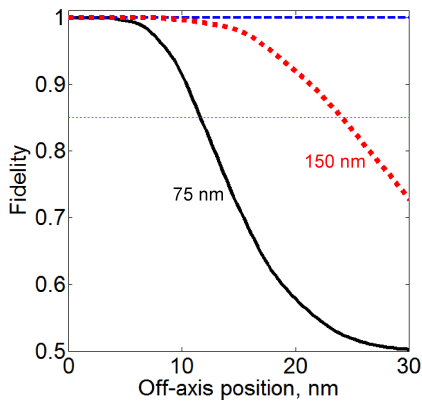}
\caption{Fidelity of entanglement for the axial (black solid and red dotted lines) and diagonal (blue dashed line) off-axis QD locations at $\lambda=1550$ nm, for two cross gap dimensions $g$ of $75$ nm and $150$ nm. The $0.85$ threshold is indicated.}
\end{figure}

In conclusion, we have demonstrated \corrected{a novel source of polarization-entangled photon pairs based on a plasmonic nanoantenna driven by a single QD. The source combines} large broadband SE enhancement and {\color{red}very}-high extraction efficiency, \corrected{which is not achievable with competing devices offering either high SE enhancement or high efficiency}. It extends the applicability of plasmonic nanoantennas to sources of nonclassical light and anticipates their integration with photonic circuits for application in quantum communication and computing.  

This work was supported by the Australian Research Council. The authors confirm invaluable discussions with I. Staude, P. Belov, and F. Demming (CST). 

\providecommand{\noopsort}[1]{}\providecommand{\singleletter}[1]{#1}%

\end{document}